# Effort-Oriented Classification Matrix of Web Service Composition

Zheng Li
NICTA and UNSW
School of CSE
Sydney, Australia
zheng.li@nicta.com.au

Liam O'Brien
NICTA and ANU
Research School of ISE
Canberra, Australia
liam.obrien@nicta.com.au

Jacky Keung
NICTA and UNSW
School of CSE
Sydney, Australia
jacky.keung@nicta.com.au

Xiwei Xu
NICTA and UNSW
School of CSE
Sydney, Australia
xiwei.xu@nicta.com.au

*Abstract*—Within the service-oriented computing domain, Web service composition is an effective realization to satisfy the rapidly changing requirements of business. Therefore, the research into Web service composition has unfolded broadly. Since examining all of the related work in this area becomes a mission next to impossible, the classification of composition approaches can be used to facilitate multiple research tasks. However, the current attempts to classify Web service composition do not have clear objectives. Furthermore, the contexts and technologies of composition approaches are confused in the existing classifications. This paper proposes an effort-oriented classification matrix for Web service composition, which distinguishes between the context and technology dimension. The context dimension is aimed at analyzing the environment influence on the effort of Web service composition, while the technology dimension focuses on the technique influence on the effort. Consequently, besides the traditional classification benefits, this matrix can be used to build the basis of cost estimation for Web service composition in future research.

## I. INTRODUCTION

Web services have been widely accepted as the preferred standards-based way to implement Service-Oriented Architecture (SOA) in practice. Since "only when we reach the level of service composition can we realize all the benefits of SOA" [15], the research into composing Web services has grown significantly along with the increasing necessity of reusing existing resources. In the past decade, lots of work about composing Web services has been reported in the literature. As a result, it is nearly impossible to explore this research area by exhausting all the published composition approaches. However, we can inductively classify the existing Web service composition work, and thereby to facilitate the comprehension of related knowledge.

Existing classification work of Web service composition can be found in several survey papers [16, 18]. Nevertheless, these classifications are either incomplete or ambiguous, which brings many issues when using them to categorize and analyze new composition approaches. Firstly, none of the existing classifications distinguishes between the composition technologies and the composition contexts. For example, Dustdar and Schreiner [16] list model-driven approaches as a separate composition category, while combine AI planning approaches with the automatic design process and ontology environment. Secondly, the terminology is vague in some composition classifications. For example, Rao and Su [18] use "static composition" to cover those approaches having manual workflow generation, even though the component Web service selection and binding are accomplished automatically. Finally, the lack of clear classification targets is the most significant weakness of existing classification work of Web service composition. Current classification work generally surveys composition types through subjective identification without objective constrains. The resulting classification is then hardly associated with other specific research topics like software cost estimation. For example, the declarative service composition class [16] focuses on its irregular composition architecture that is almost irrelative to the composition effort and cost.

In this paper, we present a novel classification matrix aimed at the influence on the effort of Web service composition. This matrix uses clarified terminology, and differentiates the classifications between the Context and Technology dimensions. The Context dimension includes major effort related contexts that are Pattern, Semiotics, Mechanism, Design Time and Runtime. When considering different composition patterns for the same target, orchestration deals with a central mediator while choreography is a collaboration of all the participant Web services. Within the semiotic context, Semantic Web services have more descriptions than Syntactic Web services, which can facilitate service discovery and matchmaking. Mechanism context comprises SOAP-based and RESTful composition. RESTful Web service compositions are relatively lightweight compared with SOAP-based compositions. According to the manipulation procedure before generating real composite Web service, there can be manual, semi-automatic, or automatic compositions at design time. During runtime, the dynamic and static compositions are differentiated by the adaptability of Web service composition. On the other hand, the Technology dimension is divided into well defined Workflow-based, Model-driven, and AI planning technology categories. In fact, one composition approach can be classified into one technology category and some context categories at the same time. For example, the approach in [4] uses model-driven technology and is under semantic context. Therefore, the matrix is suitable to represent this kind of cross-classification.

The contributions of this research are manifold. Firstly, the complete classification matrix can help researchers explore the knowledge space in service composition domain, and help developers choose suitable techniques when composing Web services. Secondly, since this classification matrix is effort-oriented, the different technology categories and context types can be used as cost factors when estimating the cost and effort of Web service composition. It is undoubted that different type of composition technology under different context will cost differently. Finally yet importantly, new research opportunities could be identified when comparing and analyzing different composition approaches through the classification matrix.

This paper is organized as follows. Section II introduces the context-based classification through specifying every type of context. Section III presents the technology-based classification, and explicitly defines different technology categories. The conclusion is drawn, and some potential research opportunities are identified, in section IV. At the end of this paper, a small part of our work is filled in Table I as a sample of classification matrix of Web service composition.

## II. CONTEX-BASED CLASSIFICATION

The Context discussed here refers to the environment and different stages when composing Web services. Through analyzing the lifecycle of Web service composition, we have identified Pattern, Semiotics, Mechanism, Design Time, and Runtime contexts that have the most influence on composition effort.

### A. Pattern: Orchestration and Choreography

According to the cooperation fashion among component Web services, the Web service composition patterns can be distinguished between *orchestration* and *choreography*.

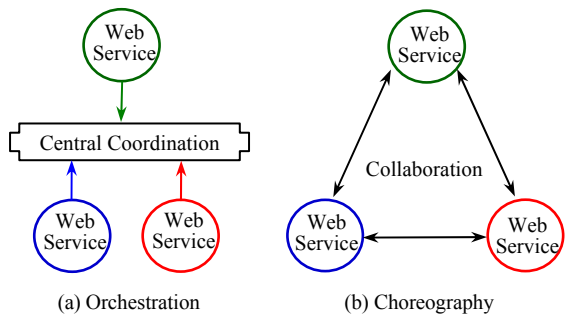

Figure 1. Web Service Orchestration and Choreography.

Orchestration, as shown in Figure 1(a), describes and executes a centralized process flow that normally acts as a coordinator to the involved Web services. The central coordinator explicitly specifies the business logic and controls the order of invocation of Web services. As a result, the coordination defines a long-term, cross-organization, transactional process. The involved Web services, on the other hand, do not need to be aware of whether they are involved in an orchestrated process. Orchestration represents coordination from the perspective of a single participant that can be another Web service.

Choreography, as shown in Figure 1(b), describes multi-party collaboration and focuses on the peer-to-peer message exchange. The collaboration is decentralized, which means that all participating Web services work equally and do not rely on a central controller. Each Web service involved in choreography knows exactly its contribution to a business process: operation, timing of operation, and the interaction with other participants. Choreography represents collaboration from a global perspective.

In most cases, the pattern to which Web service composition belongs can be identified easily through the adopted standards or flow languages. For example, the current de facto standard for Web service orchestration is the Business Process Execution Language (BPEL). BPEL is an executable business process modeling language that can be used to describe the execution logic by defining the control flow and prescribing the rules for managing the non-observable data. The BPEL engine can then execute the description and orchestrate the pre-specified activities. Whereas one of the most widespread W3C recommendation protocols for choreography is Web Services Choreography Description Language (WS-CDL). WS-CDL is designed to describe the common and collaborative observable behavior of multiple Web services that interact with each other to achieve their common goal. In other words, WS-CDL description offers the specification of collaborations between the participants involved in choreography.

### B. Semiotics: Syntactic and Semantic Composition

The semiotic environment is becoming a significant context for Web service composition along with the evolution of the Web. Semiotics is the general science of signs, which studies both human language and formal languages. *Syntax* and *semantics* are two of fundamental components of semiotics. Syntax is related to the formal or structural relations between signs and the production of new ones, while semantics deals with the relations between the sign combinations and their inherent meaning.

Currently, the World Wide Web can be mainly considered as syntactic Web that uses Hyper Text Markup Language (HTML) to compose documents and publish information. When it comes to Web services, the syntactic level XML standards, for example Web Service Description Language (WSDL) and Universal Description, Discovery and Integration (UDDI), have spread widely to address corresponding e-business activities and research issues in industry and academia. However, the syntactic Web was designed primarily for human interpretation and use, so that the inherent meaning of information on the syntactic Web is not understandable for computers and applications.

To overcome the obstacles of interpretability and interoperability between traditional systems and applications, the semantic Web was proposed through incremental and information-added adjustments. The semantic Web is not a separate Web but brings machine-understandable and human-transparent descriptions to the existing data and documents on the syntactic Web. As for semantic Web services, the information needed to select, compose, and respond to services can be encoded with semantic markup at the service Web sites. Accordingly, new standards and languages of semantic markup, like Web Ontology Language for Web Services (OWL-S) and Web Service Modeling Ontology (WSMO), should be investigated and used to give meaning to Web services. Driven by the semantic markup and agent technologies, semantic Web service discovery, selection, composition, and execution are all supposed to be automatic tasks. Although the full automation of these processes are still the objects of ongoing research, accomplishing parts of this goal can be achieved. For example, the semantic description is helpful for the translation between Web service composition problems and AI-planning systems [12], while the semantic matchmaking can facilitate the automatic Web service discovery [1].

Overall, the XML-based standards are for syntax, whilst the ontology-based ones are for semantics. Both share unified Web infrastructure and together provide capability for developing Web applications that deal with data and semantics. The most important characteristic of ontology-based techniques is that they allow a richer integration and interoperability of data among communications and domains. Consequently, Web service compositions can be categorized according to syntactic and semantic context, while the context can be also identified through employed standards and techniques.

### C. Mechanism: RESTful and SOAP-based Composition

Concentrating on the technologies and architectures, nowadays there are two main mechanism paradigms of building composite Web services, namely *RESTful* composition and *SOAP-based* composition.

Basically, REpresentational State Transfer (REST) and Simple Object Access Protocol (SOAP) are not directly comparable with each other and not necessarily opposites. REST is an architectural style originally designed for building large-scale distributed hypermedia systems, whereas SOAP is a general protocol used as one foundation of numerous WS-* technologies. Within the REST environment, the Web is considered as a universal store medium for publishing globally accessible information. In contrast, SOAP/WS-* treats the Web as the universal transport medium for exchanging messages. When building Web services, traditional SOAP/WS-* environment requires relatively heavyweight open standards than that used in RESTful context. Although the SOAP vs. REST debate has been an ongoing discussion for some time, these is an implicit consensus that REST is more suitable for basic, ad-hoc, client-driven scenarios, while SOAP/WS-* are used to address the more advanced quality of services requirements in business related high-interactive applications.

SOAP/WS-* based Web service composition is a collection of related, structured activities or tasks that produce a specific service or product for a particular customer. On the other hand, RESTful Web service composition integrates normally disparate Web resources to create a new application. These resources can be the exposure of pure data or traditional application functionality. The execution of RESTful Web service composition is based on a web browser.

### D. Classification at Different Composition stages

Generally, there are four fundamental activities when composing Web services, namely Planning, Discovery, Selection, and Execution [17]. The Planning is to determine a composition plan including the execution sequence of tasks. Every task is a service functionality or a service activity. The Discovery is to find all the candidate services that can satisfy the tasks in the plan. The aim of Selection is to choose optimal subset from all the discovered services by using non-functional attributes. Finally, the Execution will build a real composite Web service.

Meanwhile, the Web service composition process can be separated into design time and runtime stages. Figure 2 shows one of the possible composition scenarios. Depending on the real practices, the design time stage can comprise various activities from only Planning to the combination of Planning, Discovery, and Selection. Moreover, the sequence of Planning, Discovery, and Selection can also be diverse. For example, the theorem proving approach in [12] is based on the pre-determined Web services to generate the composition plan. According to the involved effort, the design time procedure can be *manual*, *semi-automatic*, and *automatic*. Considering there is still a long way to realize the complete automation of Web service composition even at design time, we only concentrate on the Planning activity when unfolding classification.

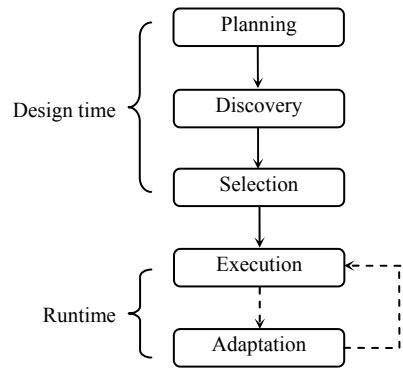

Figure 2. A Web Service Composition Scenario.

The Execution activity stays at runtime stage of Web service composition. During the execution process, the network configurations and non-functional factors may change, and existing Web services may be updated or terminated. Consequently, some pre-identified service may not be available and another tradeoff one needs to be re-selected or re-discovered. In order to adapt and even take advantage of the changing environment, there is a possible Adaptation activity during execution. Therefore, we can define that the Web service composition is *dynamic* at runtime if it is adaptive with minimal user intervention, otherwise *static*.

## III. Technology-based Classification

The Technology refers to the techniques used in the approaches to implement Web service composition. It is difficult to enumerate all kinds of composition techniques, although different technique can contribute different composition effort. Fortunately, we can identify three group of techniques: Workflow-based, Model-driven, and AI planning techniques.

### A. Workflow-based Techniques

Workflow is a virtual representation of actual work including a sequence of operations. Workflow-based Web service composition uses the workflow perspective to describe the normally complex collaboration among Web services and implement the composition procedure. There are two ways to describe the Web service composition workflow:

*1) To program the executable workflow directly:* Obviously, the composition process can be programmed from scratch by using traditional languages and standards. Whereas the current universal technique is to use the dedicated, process-oriented language like BPEL to specify the transition interactions among Web services at a macro-level state.

*2) To draw the abstract workflow without programming:* Supported by some tools or engines, the workload of Web service composition can be relieved by drawing the abstract workflow without programming. For example, the semantic matchmaking based approach [1] uses the GUI panel of composer to construct an abstract flow, while eFlow [2] adopts a graph-oriented method to define the interaction and order of execution among the nodes in an abstract composition process.

### B. Model-driven Techniques

In model-driven approaches of Web service composition, models are used to describe user requirements, information structures, abstract business processes, component services and component service interactions. The models are independent of, but can be tranformed into, executable composition specifications. Generally, there is also modeling work in some workflow-based techniques. Whereas the model-driven techniques discussed here merely follow the standards provided by the Object Management Group (OMG). The standards mainly refer to the Unified Modeling Language (UML) and Model-Driven Architecture (MDA).

Numerous discussions related to UML-based modeling of Web service composition can be found in the literature. The generic scenario is to use UML class diagrams to represent the state parts of compositions, while the behaviour parts are represented through UML activity diagrams. The state parts can be Web service interface [3], the structure of composite Web service [4], and QoS characteristics [5]. On the other hand, the behaviour parts describe the composition operations, interactions of component Web services, and control and data flow. Furthermore, since BPEL is widely accepted for composing Web services, UML has been designedly extended for BPEL to cover most aspects of Web service composition. Particularly, the UML-WSC profile [6] can be used as an alternative language to BPEL, and be executed directly through the proposed process engine.

### C. AI Planning Techniques

AI planning seeks to use intelligent systems to generate a plan that can be one possible solution to a specified problem, while a plan is an organized collection of operators within the given application domain. AI planning is essentially a search problem. The underlying basis of planning relies on state transition system with states, actions and observations. Benefiting from the state transition system, the planner explores a potentially large search space and produces a plan that is applicable to bridge the gap between the initial state and goal when run. Therefore, AI planning in Web service composition normally comprises of five attributes: all the available services, the initial state, the state change functions, all the possible states, and the final goal. The initial state and final goal are specified in the requirements for composing Web service. The state change functions define the preconditions and effects when invoking Web services.

A large amount of research has been reported about the AI planning related Web service composition. These works apply techniques ranging from Situation Calculus [7], Automata Theory [8], Rule-based Planning [9], Query Planning [11], Theorem Proving [12], Petri Nets [13], to Model Checking [14]. Generally, these techniques convert the problems of composition into generating execution workflows using the respectively dedicated expression. The workflows can then be transformed into executable specifications like BPEL documents or other XML-based descriptions, and executed through the corresponding engines.

## IV. Conclusion

The territory of Web service composition has been researched so broadly that it becomes difficult to explore every existing composition approach. However, we can deliver a general classification of Web service composition through investigating limited approaches inductively. Unlike existing classification work, this paper proposes an effort-oriented classification matrix of Web service composition. The matrix uses two dimensions, Context and Technology, to classify different compositions. Several pairs of effort-related contexts are picked in the Context dimension, while three technique categories are paralleled in the Technology dimension. This effort-oriented classification matrix can be used to facilitate exploration and comprehension in the research area of Web service composition, cost and effort estimation for compositing Web services, and identification of new research opportunities. In fact, based on our current work, some new research opportunities in Web service composition area have been already revealed. For example bridging the gap between automatic composition at design time and dynamic composition at runtime.

Overall, the work described in this paper brings a new perspective of classification of Web service composition. The prominent characteristic of the proposed classification matrix is the specific objective - aiming at the influence on the effort of Web service composition. Our future work is to continue filling this classification matrix, and to use the matrix to build the basis of cost estimation for Web service composition.

ACKNOWLEDGMENT

NICTA is funded by the Australian Government as represented by the Department of Broadband, Communications and the Digital Economy and the Australian Research Council through the ICT Centre of Excellence program.

REFERENCES

[1] E. Sirin, J. Hendler, and B. Parsia, "Semi-automatic Composition of Web Services using Semantic Descriptions," Web Services: Modeling, Architecture and Infrastructure workshop in ICEIS 2003, Apr. 2003.

[2] F. Casati, S. Ilnicki, L. Jin, V. Krishnamoorthy, and M. Shan, "Adaptive and Dynamic Service Composition in EFlow," Proc. 12th International Conference on Advanced Information Systems Engineering (CaiSE*00), Springer, Jun. 2000, pp. 13-31, doi: 10.1007/3-540-45140-4_3.

[3] D. Skogan, R. Groenmo, and I. Solheim, "Web Service Composition in UML," Proc. Eighth IEEE International Enterprise Distributed Object Computing Conference (EDOC 2004), IEEE Press, Sept. 2004, pp. 47-57, doi: 10.1109/EDOC.2004.1342504.

[4] J. T. E. Timm and G. C. Gannod, "Specifying Semantic Web Service Compositions using UML and OCL," Proc. 2007 IEEE International Conference on Web Services (ICWS 2007), IEEE Press, Jul. 2007, pp. 521-528, doi: 10.1109/ICWS.2007.168.

[5] R. Grønmo and M. C. Jaeger, "Model-driven Semantic Web Service Composition," Proc. 12th Asia-Pacific Software Engineering Conference (APSEC '05), IEEE Press, Dec. 2005, pp. 15-17, doi: 10.1109/APSEC.2005.81.

[6] S. Thone, R. Depke, and G. Engels, "Process-Oriented, Flexible Composition of Web Services with UML", Proc. Third International Joint Workshop on Conceptual Modeling Approaches for E-business: A Web Service Perspective (eCOMO 2002), Springer, Oct. 2002, pp. 390-401, doi: 10.1007/b12013.

[7] V. R. Chifu, I. Salomie, and E. St. Chifu, "Fluent Calculus-based Web Service Composition — From OWL-S to Fluent Calculus," Proc. 4th International Conference on Intelligent Computer Communication and Processing (ICCP 2008), IEEE Press, Aug. 2008, pp. 161-168, doi: 10.1109/ICCP.2008.4648368.

[8] S. Mitra, R. Kumar, and S. Basu, "Automated Choreographer Synthesis for Web Services Composition Using I/O Automata," Proc. IEEE International Conference on Web Services (ICWS 2007), IEEE Press, Jul. 2007, pp. 364-371, doi: 10.1109/ICWS.2007.47.

[9] B. Medjahed, A. Bouguettaya, and A. K. Elmagarmid, "Composing Web services on the Semantic Web", The VLDB Journal, vol. 12, Sept. 2003, pp. 333-351, doi: 10.1007/s00778-003-0101-5.

[10] S. V. Hashemian and F. Mavaddat, "A Graph-based Approach to Web Services Composition," Proc. The 2005 Symposium on Applications and the Internet, IEEE Press, Jan.-Feb. 2005, pp. 183-189, doi: 10.1109/SAINT.2005.4.

[11] S. Thakkar, C. Knoblock, and J. Ambite. "A View Integration Approach to Dynamic Composition of Web Services," Proc. 2003 ICAPS Workshop on Planning for Web Services, AAAI Press, Jul. 2003.

[12] J. Rao, P. Küngas, and M. Matskin, "Composition of Semantic Web Services using Linear Logic Theorem Proving," Information Systems, vol. 31, Jun.-Jul. 2006, pp. 340-360, doi: 10.1016/j.is.2005.02.005.

[13] V. Gehlot and K. Edupuganti, "Use of Colored Petri Nets to Model, Analyze, and Evaluate Service Composition and Orchestration," Proc. 42nd Hawaii International Conference on System Sciences (HICSS'09), IEEE Press, Jan. 2009, pp. 1-8, doi: 10.1109/HICSS.2009.487.

[14] P. Traverso and M. Pistore, "Automated Composition of Semantic Web Services into Executable Processes," Proc. Third International Semantic Web Conference (ISWC'04), Nov. 2004, pp. 380-394.

[15] P. Sarang, F. Jennings, M. Juric, and R. Loganathan, SOA Approach to Integration: XML, Web services, ESB, and BPEL in real-world SOA projects. Packt Publishing, Birmingham, 2007.

[16] S. Dustdar and W. Schreiner, "A Survey on Web Services Composition," International Journal of Web and Grid Services, vol. 1, Aug. 2005, pp. 1-30, doi: 10.1504/IJWGS.2005.007545.

[17] J. Cardoso and A. P. Sheth, Semantic Web Services, Processes and Applications. New York: Springer, 2006.

[18] J. Rao and X. Su, "A Survey of Automated Web Service Composition Methods," Lecture Notes in Computer Science, vol. 3387/2005, Jan. 2005, pp. 43-54, doi: 10.1007/b105145.

[19] F. Rosenberg, F. Curbera, M. J. Duftler, and R. Khalaf, "Composing RESTful Services and Collaborative Workflows: A Lightweight Approach," IEEE Internet Computing, vol. 12, Sept.-Oct. 2008, pp. 24-31, doi: 10.1109/MIC.2008.98.

[20] J. Lathem, K. Gomadam, and A. P. Sheth, "SA-REST and (S)mashups : Adding Semantics to RESTful Services," Proc. First IEEE International Conference on Semantic Computing (ICSC 2007), IEEE Press, Sept. 2007, pp. 469-476, doi: 10.1109/ICSC.2007.94.

[21] R. Anzboeck and S. Dustdar, "Semi-Automatic Generation of Web Services and BPEL Processes - A Model-Driven Approach," Lecture Notes in Computer Science, vol. 3649/2005, Sept. 2005, pp. 64-79, doi: 10.1007/11538394_5.

[22] V. Valero, M. E. Cambronero, G. Díaz, and H. Macià, "A Petri Net Approach for the Design and Analysis of Web Services Choreographies," Journal of Logic and Algebraic Programming, vol. 78, May-Jun. 2009, pp. 359-380, doi: 10.1016/j.jlap.2008.09.002.

[23] S. G. H. Tabatabaei, W. M. N. Kadir, and S. Ibrahim, "Semantic Web Service Discovery and Composition Based on AI Planning and Web Service Modeling Ontology," Proc. IEEE Asia-Pacific Services Computing Conference (APSCC '08), IEEE Press, Dec. 2008, pp. 397-403, doi: 10.1109/APSCC.2008.126.

[24] E. Sirin, B. Parsia, D. Wu, J. Hendler, and D. Nau, "HTN Planning for Web Service Composition using SHOP2," Web Semantics: Science, Services and Agents on the World Wide Web, vol. 1, Oct. 2004, pp. 377-396, doi: 10.1016/j.websem.2004.06.005.

[25] H. Zhao and P. Doshi, "Towards Automated RESTful Web Service Composition," Proc. IEEE International Conference on Web Services (ICWS 2009), IEEE Press, Jul. 2009, pp. 189-196, doi: 10.1109/ICWS.2009.111.

[26] F. Casati, M. Sayal, and M. Shan, "Developing E-Services for Composing E-Services," Lecture Notes in Computer Science, vol. 2068/2001, Jan. 2001, pp. 171-186, doi: 10.1007/3-540-45341-5_12.

[27] X. Xu, L. Zhu, Y. Liu, and M. Staples, "Resource-Oriented Architecture for Business Processes," Proc. 15th Asia Pacific Software Engineering Conference (APSEC2008), IEEE Press, Dec. 2008, pp. 395-402, doi: 10.1109/APSEC.2008.52.

[28] S. Mosser, "Web Services Composition: Mashups Driven Orchestration Definition," Proc. 2008 Int'l Conf. Computational Intelligence for Modeling Control & Automation, IEEE Press, Dec. 2008, pp. 284-289, doi: 10.1109/CIMCA.2008.96.

[29] Y. Xu, S. Tang, Y. Xu, and Z. Tang, "Towards Aspect Oriented Web Service Composition with UML," Proc. 6th Int'l. Conf. Computer and Information Science (ICIS2007), IEEE Press, Jun. 2007, pp. 279-284, doi: 10.1109/ICIS.2007.185.

[30] J. Pathak, S. Basu, R. Lutz, and V. Honavar, "MoSCoE: A Framework for Modeling Web Service Composition and Execution," Proc. 22nd International Conference on Data Engineering Workshops, IEEE Press, Apr. 2006, pp. x143, doi: 10.1109/ICDEW.2006.96.

TABLE I.  A SAMPLE OF CLASSIFICATION MATRIX OF WEB SERVICE COMPOSITION

| Technology | | Context | | | | | | | | | |
|---|---|---|---|---|---|---|---|---|---|---|---|
| | | *Pattern* | | *Semiotics* | | *Mechanism* | | *Design Time* | | | *Runtime* |
| *Category* | *Detailed Technique* | *Orchestration* | *Choreography* | *Syntax* | *Semantics* | *SOAP* | *REST* | *Manual* | *Semi-Auto* | *Auto* | *Static* | *Dynamic* |
| Workflow-based | BPEL Programming | √ | | √ | | √ | | √ | | | √ | |
| | Semantic Matching [1] | | √ | | √ | √ | | | √ | | √ | |
| | eFlow [2] | √ | | √ | | √ | | √ | | | | √ |
| | Bite [19] | | √ | √ | | | √ | √ | | | √ | |
| | SA-REST + Smashup [20] | √ | | | √ | | √ | | √ | | √ | |
| | CSDL [26] | √ | | √ | | √ | | √ | | | √ | |
| | RESTfulBP [27] | | √ | √ | | | √ | √ | | | √ | |
| Model-driven | UML + MDA [3] | √ | | √ | | √ | | √ | | | √ | |
| | UML + OCL [4] | √ | | | √ | √ | | √ | | | | √ |
| | UML + QoS Support [5] | √ | | | √ | √ | | | √ | | √ | |
| | UML-WSC [6] | √ | | √ | | √ | | √ | | | | √ |
| | UML + IHE framework [21] | √ | | √ | | √ | | √ | | | | √ |
| | MD Mashup [28] | | √ | √ | | | √ | √ | | | √ | |
| | UML-AOWSC [29] | √ | | √ | | √ | | √ | | | | √ |
| | MoSCoE [30] | √ | | | √ | √ | | | √ | | √ | |
| AI planning | SHOP2 [24] | √ | | | √ | √ | | | | √ | √ | |
| | Petri Net [22] | | √ | | √ | √ | | | | √ | √ | |
| | Situation Calculus [7] | √ | | | √ | √ | | | | √ | √ | |
| | I/O Automata [8] * | | √ | √ | | √ | | | | √ | √ | |
| | Rule-based Planning [9] | √ | | | √ | √ | | | | √ | √ | |
| | Interface Automata [10] | √ | | | √ | √ | | | | √ | √ | |
| | Query Planning [11] * | √ | | √ | | √ | | | | √ | √ | |
| | Linear Logic Theorem Proving [12] | √ | | | √ | √ | | | √ | | √ | |
| | Colored Petri Net [13] | √ | | √ | | √ | | | | √ | √ | |
| | Model Checking [14] | √ | | | √ | √ | | | | √ | √ | |
| | AIMO [23] | | √ | | √ | √ | | | | √ | | √ |
| | Situation Calculus for REST [25] | √ | | | √ | | √ | √ | | | √ | |

* The approaches in [8] and [11] are independent of the Semiotics context.